\documentclass[twocolumn,9 pt]{extarticle}
\usepackage{amsmath,amssymb}
\usepackage[top=0.85in,left=0.5in, right=0.5in, footskip=0.5in]{geometry}
\usepackage{cite}
\usepackage{nameref,hyperref}
\usepackage{graphicx}
\usepackage{fixltx2e}
\usepackage{authblk}
\begin{document}

\title{\textsc{Imaging Renal Urea Handling in Rats at Millimeter Resolution using Hyperpolarized Magnetic Resonance Relaxometry}}

\author[1,2]{Galen D. Reed
	\thanks{Corresponding author email: galen.d.reed@gmail.com}}
\author[1,2]{Cornelius von Morze}
\author[3]{Alan S. Verkman}
\author[1,2]{Bertram L. Koelsch}
\author[1]{Myriam M. Chaumeil}
\author[4]{Michael Lustig}
\author[1,2]{Sabrina M. Ronen}
\author[5]{Jeff M. Sands}
\author[1,2]{Peder E. Z. Larson}
\author[1]{Zhen J. Wang}
\author[6,7]{Jan Henrik Ardenkj\ae r Larsen}
\author[1,2]{John Kurhanewicz}
\author[1,2]{Daniel B. Vigneron}

\affil[1]{Department of Radiology and Biomedical Imaging, University of California San Francisco, San Francisco, California, USA}
\affil[2]{Graduate Group in Bioengineering, University of California San Francisco, San Francisco, California, USA, and 
University of California Berkeley, Berkeley, California, USA }
\affil[3]{Departments of Medicine and Physiology, University of California San Francisco, San Francisco, California, USA }
\affil[4]{Department of Electrical Engineering and Computer Sciences, University of California Berkeley, Berkeley, California, USA}
\affil[5]{Department of Medicine, Renal Division, Emory University, Atlanta, Georgia, USA }
\affil[6]{GE Healthcare, Br{\o}ndby, Denmark }
\affil[7]{Department of Electrical Engineering, Technical University of Denmark, Kongens Lyngby, Denmark }

\date{}
\maketitle

\abstract{

\textit{In vivo} spin spin relaxation time ($T_2$) heterogeneity of hyperpolarized \textsuperscript{13}C urea in the rat kidney was investigated. Selective quenching of the vascular hyperpolarized \textsuperscript{13}C signal with a macromolecular relaxation agent revealed that a long-$T_2$ component of the \textsuperscript{13}C urea signal originated from the renal extravascular space, thus allowing the vascular and renal filtrate contrast agent pools of the \textsuperscript{13}C urea to be distinguished via multi-exponential analysis. The $T_2$ response to induced diuresis and antidiuresis was performed with two imaging agents: hyperpolarized \textsuperscript{13}C urea and a control agent hyperpolarized bis-1,1-(hydroxymethyl)-1-\textsuperscript{13}C-cyclopropane-$^2\textrm{H}_8$. Large $T_2$ increases in the inner-medullar and papilla were observed with the former agent and not the latter during antidiuresis suggesting that $T_2$ relaxometry may be used to monitor the inner-medullary urea transporter (UT)-A1 and UT-A3 mediated urea concentrating process. Two high resolution imaging techniques - multiple echo time averaging and ultra-long echo time sub-2 mm$^3$ resolution 3D imaging - were developed to exploit the particularly long relaxation times observed.

\section*{Introduction}

Urea is primary end product of nitrogen metabolism in mammals, and humans typically produce more than 20 g of urea per day \cite{guyton_book}. Concentrating urea to more than 100 times plasma levels for efficient excretion of this large osmotic load while minimizing water loss is one of the primary functions of the mammalian kidney \cite{marsh_urea_review,bankir_urea_handling,fenton_review,sands_urea_regulation}. Renal urea handling is a multistep process beginning with filtration of the blood at the glomerulus followed by a countercurrent multiplication in the medulla. The countercurrent exchange is assisted by urea transporters (UT) expressed in the descending thin limb of the loops of Henle (UT-A2 isoform), the erythrocytes (UT-B isoform), and the inner two thirds of the inner medullary collecting ducts (UT-A1 and UT-A3 isoforms) \cite{fenton_review, sands_urea_regulation,yao_utb,esteva_uta}. The inner medullary collecting duct (IMDC) transporters UT-A1 and UT-A3 increase the effective permeability of the tubular wall in the presence of vasopressin. This allows for urea to freely pass into the inner medullary interstitial fluid (IF) where it can accumulate to greater than 1 M concentration when water conservation is important. 



Imaging renal solute handling could be a potentially valuable tool for the study of renal function. Several magnetic resonance imaging (MRI) studies have demonstrated non-invasive sodium detection with \textsuperscript{23}Na-detected MRI \cite{maril_namri,maril_namri_human}. Indirect urea detection via changes in the \textsuperscript{1}H water resonance after radio frequency (RF) saturation of the urea amide frequency has been demonstrated \cite{dagher_urea}. Direct magnetic resonance imaging (MRI) of intravenously-injected \textsuperscript{13}C-labeled urea \cite{GolmanEndogenous,morze_ut,patrick_urea_transgene,pages_urea,morze_urea} and numerous other small molecules \cite{GolmanMetabolic,GolmanMetabolicCancer,keshari_review,clatworthy_fumarate,grant_diffusible,SvenssonC13SSFP,merritt_pnas,laustsen_oxygen,keshari_dha,keshari_fructose,keshari_ligand} has recently been enabled using dynamic nuclear polarization (DNP) \cite{ardenkjaer_pnas}. In this process, the isotopically-enriched molecule is doped with an organic radical, cooled to liquid helium temperatures, and irradiated at microwave frequencies to achieve polarizations many orders of magnitude above thermal equilibrium values. After dissolution, the polarization decays exponentially with a time constant  $T_1$ which is typically on the order of 10-90 s for \textsuperscript{13}C-labeled carbonyl sites \cite{chattergoon_t1,keshari_review}. This method has been utilized to generate background-free angiograms in preclinical models using highly biocompatible contrast agents \cite{GolmanEndogenous,golman_agio}.

Recently, a method for \textit{in vivo} imaging of hyperpolarized \textsuperscript{13}C urea with improved spatial resolution was presented which used a $T_2$-weighted steady state free precession (SSFP) sequence in combination with supplementary \textsuperscript{15}N labeling of the urea amide groups \cite{reed_n15urea}. $T_2$ mapping experiments performed in Sprague Dawley rats gave uniform urea $T_2$ values of approximately 1 s throughout the animal with the exception of the kidneys, where values between 4 s and 15 s were noted \cite{reed_n15urea}. Given the importance if urea in renal solute handling process as well as the paucity of literature discussing the physiological significance of \textit{in vivo} \textsuperscript{13}C $T_2$ contrast, this current study presents measurements for further investigation of the \textsuperscript{13}C urea $T_2$ heterogeneity. In the first set of experiments, a chase infusion of an intravascular macromolecular relaxation agent was performed after hyperpolarized \textsuperscript{13}C urea injection but prior to $T_2$ measurement. This experiment was designed to selectively quench the vascular \textsuperscript{13}C urea signal while monitoring its effects on the \textsuperscript{13}C $T_2$ signal decay. A multi-exponential fitting algorithm tracked the changes in signal intensity of the short- and long-$T_2$ components both with and without the macromolecular relaxation agent. In a second set of experiments, hyperpolarized \textsuperscript{13}C $T_2$ mapping experiments were performed on rats on induced antidiuresis and osmotic diuresis with two imaging agents: hyperpolarized \textsuperscript{13}C urea and hyperpolarized bis-1,1-(hydroxymethyl)-1-\textsuperscript{13}C-cyclopropane-$^2\textrm{H}_8$ (commonly abbreviated as HMCP or HP001). In the antidiuresis state, the kidneys are in water conservation mode, and the inner medullary transporters UT-A1 and UT-A3 aid to maximally concentrate urea in the inner medulla.



\section*{Materials and Methods}

 \subsection*{Animal Handling}
Animal studies were performed under a protocol approved by the University of California San Francisco Institutional Animal Care and Utilization Committee (IACUC). Sprague Dawley rats (mean mass 400 g) were anesthetized with a 1.7\% isofluorane / oxygen mixture under a constant flow rate of 1 liter per minute. Animals were imaged in the supine position inside the birdcage coil and thermally insulated via heat pad. Contrast agents were injected via lateral tail vain catheters. Rats were housed three per cage at the UCSF Laboratory Animal Resource Center (LARC). Experiments were conducted between the hours of 5 p.m. and 1 a.m.

\begin{figure}[!h]
	\centering
	\includegraphics[width=3.7in]{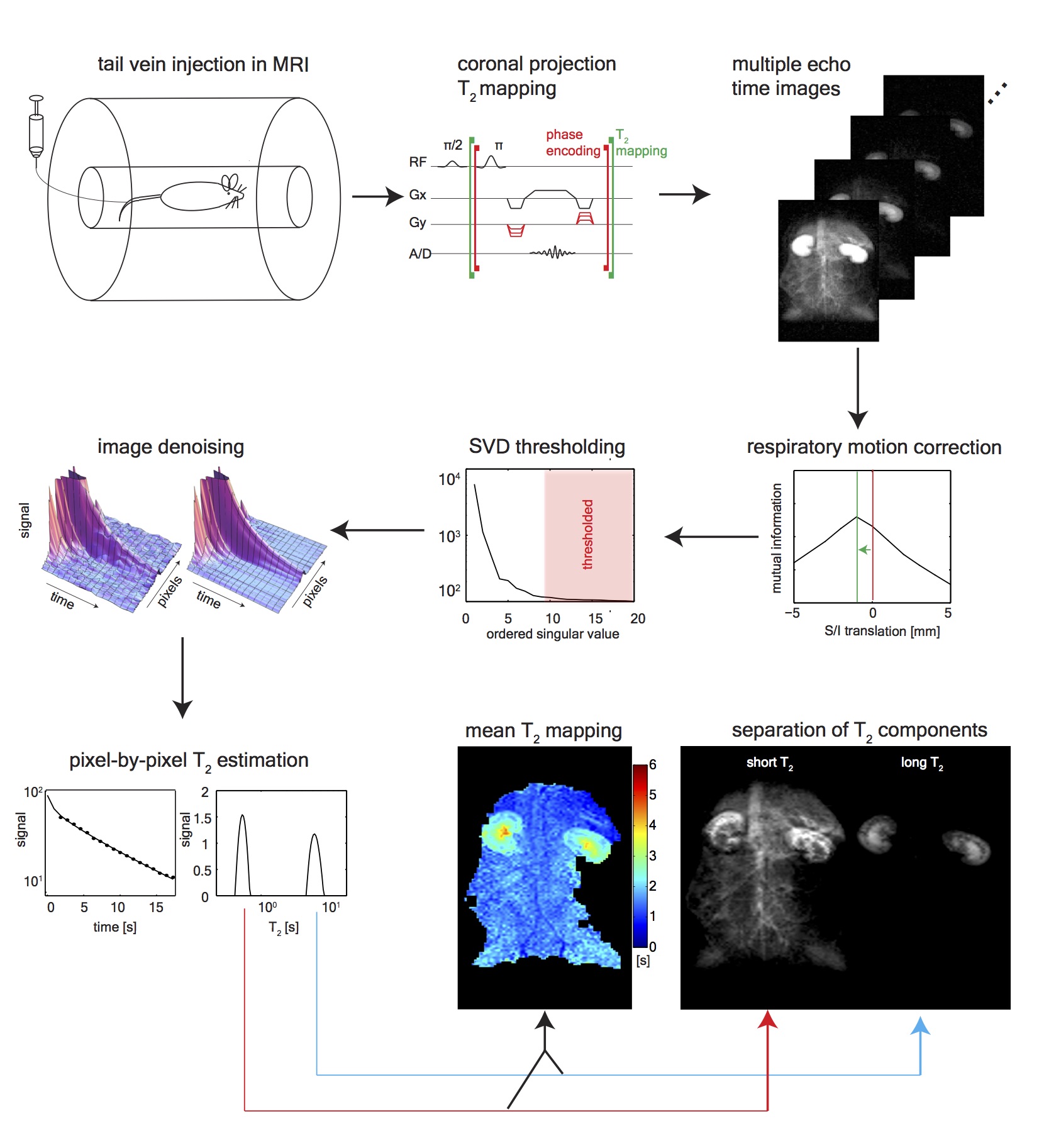}
\caption{{\bf The hyperpolarized \textsuperscript{13}C $T_2$ mapping experiment.} Hyperpolarized  \textsuperscript{13}C-labeled substrates were injected via lateral tail vain catheter inside the MRI scanner. The $T_2$ mapping sequence acquired coronal projection images at 0.9 s echo time intervals while playing 180$^\circ$ pulses  for 18 s. Each image was then corrected for respiratory motion via rigid translation in the superior / inferior direction. The dynamic images were then denoised using an SVD-based thresholding in the space / time dimensions. The signal component at each $T_2$ was estimated using a regularized version of the $T_2$ non-negative least squares method, and the long and short $T_2$ signal components were isolated by integrating the $T_2$ distribution.  
}
\end{figure}

\subsection*{Hardware}

Imaging experiments were conducted in a GE 3T clinical MRI (GE Medical Systems, Waukesha, WI) equipped with a rat-sized dual-tuned \textsuperscript{1}H / \textsuperscript{13}C transceiver birdcage RF coil (8 cm inner diameter) placed on the patient table. An Oxford Instruments Hypersense polarizer (Oxford Instruments, Oxford, UK) was used for dissolution DNP experiments.

\subsection*{Sample Preparation} 

Isotopically enriched [\textsuperscript{13}C,\textsuperscript{15}N$_2$]urea and bis-1,1-(hydroxy- methyl)-1-\textsuperscript{13}C-cyclopropane-$^2\textrm{H}_8$ were each doped with the trityl radical OX063 (Oxford Instruments, Abingdon, UK) and Dotarem (Guerbet, Roissy, France) as described previously \cite{reed_n15urea,SvenssonC13SSFP,morze_perfusion}. Supplementary urea \textsuperscript{15}N labeling  was necessary for the $T_2$ increase afforded by the elimination of the scalar coupling of the second kind relaxation pathway \cite{chiavazza_n15urea,reed_n15urea}. Bovine serum albumin (BSA) conjugated with an average of 23 gadolinium / diethylene triamine pentaacetic acid  (GdDTPA) chelates per BSA molecule (abbreviated BSA-GdDTPA, molecular weight $\sim$ 85 kDa) was synthesized using methods previously described \cite{ogan_albumin,dafni_hifmyc}.

\subsection*{\textsuperscript{13}C MRI Acquisition}
\paragraph*{$T_2$ mapping}
\textsuperscript{13}C $T_2$ mapping was performed using sequences previously described \cite{reed_n15urea}. Dynamic projection images were acquired in the coronal plane with 1 mm in-plane resolution, $14\times7$ cm $FOV$, 13 ms $TR$, and 70 phase encodes per image giving a temporal resolution of 910 ms and 18.2 s total acquisition time for all 20 echoes. Images were reconstructed with a simple 2D Fourier transform without spatial filtering. 

\subsection*{Relaxometric Data Analysis}
\paragraph*{Motion correction}
Periodic respiratory motion caused a 1-2 mm offset which was largely resolved along the superior / inferior (SI) axis of the animal. This motion can be seen in the supplementary videos S1 and S2. To correct for this observed shift, a simple search algorithm was developed in which each image was aligned with its previous time point. Given that the motion was primarily 1D, a brute-force search was implemented which translated each image in 1 mm increments over $\pm 1$ cm from the initial location along the SI axis (for a total of 20 sampling points). At each position, the normalized mutual information (MI) was calculated between the floating image and the previous time point thus generating an MI versus translation curve as shown in Fig 1. The shift which maximized this curve was then applied. 

\paragraph*{Subspace Denoising}
Outside of the kidneys, the majority of the \textsuperscript{13}C signal disappeared after the first few echoes (Fig 1). This led to low rank data matrices in the spatiotemporal dimensions. A singular value decomposition (SVD)-based denoising \cite{jha_svd_denoise} could therefore be utilized to better condition the ill-posed problem of multi-exponential estimation \cite{provencher_eigenfunction,whittall_nnls,graham_t2_criteria,raj_relaxometry}. The 3-dimensional images ($x,y,t$) were concatenated along the $x$ and $y$ dimensions forming the matrix $I(r,t)$. Singular values less than the largest 7 (out of 20) were set to zero. This SVD-thresholded matrix was then used for the regularized $T_2$ estimation (Fig 1).  

\paragraph*{Multi-Exponential Analysis}
The $T_2$ non-negative least squares ($T_2$NNLS) algorithm \cite{whittall_nnls} was used for quantitative analysis of the time-decay data. This method uses a least-squares inversion with non-negativity constraints to estimate $s(T_2)$, the signal's $T_2$ distribution at each pixel. $T_2$NNLS is appealing since it requires no \textit{a priori} assumptions on the number of decay-modes. However, regularization, necessary for management of noise amplification, imposes assumptions of the shape of the distribution. $L_2$ norm regularization produces smoothly varying distributions, whereas $L_1$ norm regularization generates a sparse $T_2$ spectra \cite{whittall_nnls}. We observed similar image appearance when using $L_1$- and $L_2$-regularized inversion of separation of short- and long-$T_2$ signal components, but the sparse $T_2$ spectra was easier to interpret visually (for example, see Fig. 3d). The inverse problem with $L_1$ constraints minimized       
\begin{equation}
||As - y||^2 + \lambda ||s||_1, 
\end{equation}
where the matrix $A$ has the elements
\begin{equation}
A_{i,j} = e^{-t_j/T_{2,i}}. 
\end{equation}
  $T_{2,i}$ is an array of 128 logarithmically-spaced $T_2$ values ranging from .3 to 20 s, $t_j$ are the echo times (20 regularly spaced from 0.5 to 19 s), and $y_j$ is the detected pixel $SNR$ values of the $j$th echo. The regularization parameter $\lambda$ was chosen to be 0.01\% of $\lambda_{\textrm{max}}$, where $\lambda_{\textrm{max}}$ the maximum possible value for which the solution is non-zero: $\lambda_{\textrm{max}} =  2\textrm{ max} \left(A^T y\right)$  \cite{kim_l1ls}. 
  
From the distribution $s(T_2)$, the signal $s_{\alpha,\beta}$ in the range $T_2\in [\alpha,\beta]$ was calculated by integrating the $s(T_2)$ from $\alpha$ to $\beta$. A mean $\langle T_2\rangle$ measure in the range $T_2\in [\alpha,\beta]$ was estimated by the first moment of the distribution:
\begin{equation}\label{eq_meant2}
\langle T_2\rangle =  \exp\left\{\frac{\int_\alpha^\beta\! s(T_2)\log T_2\,dT_2}{\int_\alpha^\beta\! s(T_2)\,dT_2}\right\}.  
\end{equation}
The integration limits $\alpha$ and $\beta$ were used to isolate short- or long-$T_2$ species.

\subsection*{\textsuperscript{13}C Urea / BSA-GdDTPA Relaxometry}
\paragraph{$^{13}$C Imaging Experiments}
These experiments were designed to selectively quench the vascular \textsuperscript{13}C urea polarization while measuring the differential attenuation of the short and long $T_2$ signal components. Analogous to previously reported experiments \cite{smith_gdchaser}, a gadolinium (Gd) carrier molecule was infused after the hyperpolarized \textsuperscript{13}C injection but prior to \textsuperscript{13}C detection. The paramagnetic Gd greatly reduces the $T_1$ of the \textsuperscript{13}C molecule thus causing rapid \textsuperscript{13}C polarization loss in regions where the \textsuperscript{13}C-labeled molecule and the Gd carrier are in close proximity. This polarization loss manifests as lowered signal during \textsuperscript{13}C imaging. This study used bovine serum albumin as the carrier which was conjugated with and average of 23 Gd / diethylene triamine pentaacetic acid chelates (abbreviated BSA-GdDTPA). This a well-characterized blood pool agent has a high molecular weight (85 kDa) which prohibits extravasation and glomerular filtration on the sub minute timescale \cite{ogan_albumin,dafni_hifmyc}. 

To estimate the efficacy of hyperpolarized \textsuperscript{13}C signal quenching, the \textsuperscript{13}C $T_1$ relaxivity of [\textsuperscript{13}C,\textsuperscript{15}N$_2$]urea with respect to BSA-GdDTPA was measured. BSA-GdDTPA was titrated in approximately 0.2 mM increments into a into a 1 mL vial containing 1 M [\textsuperscript{13}C,\textsuperscript{15}N$_2$]urea. At each titration point, the \textsuperscript{13}C urea $T_1$ was measured via saturation recovery experiments at $B_0=3$ T, $T=27^\circ$C. A linear fit was performed on the [BSA-GdDTPA] versus $1/T_1$ data points. Using this measured curve in conjunction with model-based estimates of the rats' blood volumes, the \textit{in vivo} $T_1$ of \textsuperscript{13}C urea within the blood pool was roughly estimated.

In this study, 4 rats were imaged after 2 separate \textsuperscript{13}C urea injections: one with the BSA-GdDTPA chaser and one without to act as the control experiment. Due to the persistence of the BSA-GdDTPA in the blood pool, the chaser experiment was always performed after the control. Fig 3a shows a schematic of the experimental timeline. Rats were injected with 3 mL, $150$ mM hyperpolarized [\textsuperscript{13}C,\textsuperscript{15}N$_2$]urea solution over 12 s. The hyperpolarized urea was then allowed to diffuse for 28 s, and the \textsuperscript{13}C $T_2$ mapping sequence was initiated 40 s after the beginning of injection. 2 hours later, a second [\textsuperscript{13}C,\textsuperscript{15}N$_2$]urea infusion was performed over 12 s. The hyperpolarized urea was then allowed to diffuse for 20 s, then 1 mL, $0.59$ mM BSA-GdDTPA (15 mM GdDTPA) was injected over 1 s. The chaser was allowed to diffuse for 7 s, and then the \textsuperscript{13}C $T_2$ mapping sequence was initiated again 40 s after the beginning of the urea injection.

\paragraph{Statistical Analysis}
Image noise was estimated by the standard deviation of a signal-free region of the \textsuperscript{13}C images (prior to image denoising), and all image pixels were normalized by this measurement. All pixels of the \textsuperscript{13}C urea images with first-time-point-SNR greater than 10 were included in analysis. Multi exponential calculations were performed on each pixel after image alignment and denoising. The short- and long-$T_2$ signal components were isolated from the $T_2$ distributions calculated at each pixel using the integration limits ($\alpha = 0.3$ s, $\beta = 2.5$ s) for the short-$T_2$ component and ($\alpha =2.5$ s, $\beta= 20$ s) for the long-$T_2$ component. In this way, 4 maps were calculated for each animal: short- and long-$T_2$ signal maps for each \textsuperscript{13}C urea and \textsuperscript{13}C urea + BSA-GdDTPA experiment. In each animal, the mean pixel signal of each map was computed, and two paired t-tests were performed: one comparing the mean short-$T_2$ SNR of the \textsuperscript{13}C urea images with and without BSA-GdDTPA chaser, and the other comparing the mean long-$T_2$ SNR of the \textsuperscript{13}C urea images with and without BSA-GdDTPA chaser. $p<0.05$ was used as the significance criteria. Additionally, all pixels from the short-$T_2$ and long-$T_2$ maps from all 4 animals were binned by signal intensity and plotted on top of each other as a aid for visualization of the BSA-GdDTPA chaser effects. 

\paragraph{$^1$H MRI}
For a qualitative visualization of the Gd carrier distribution, \textsuperscript{1}H-detected, $T_1$-weighted images were acquired 5 minutes post infusion of BSA-GdDTPA. This was followed by an additional injection of 1 mL, $0.5$ mM low molecular weight (940 Da) Gd-DTPA without attached albumin (Magnevist, Bayer Schering, Berlin). Both Gd agents were imaged the same \textsuperscript{1}H spoiled gradient echo (SPGR) sequence (flip angle = 35, $TE/TR=1.4/7$ ms, 3 averages, .8 mm isotropic resolution).

 \subsection*{Diuresis / Antidiuresis Relaxometry}
\paragraph{$^{13}$C Imaging Experiments}

These experiments were designed to measure the $T_2$ of two agents - hyperpolarized \textsuperscript{13}C urea and hyperpolarized \textsuperscript{13}C HMCP - in response to induced antidiuresis and osmotic diuresis in rats. In the antidiuresis state, the kidney is in maximally concentrating mode, and high levels of circulating vasopressin activate the inner medullary transporters UT-A1 and UT-A3 in order to maximally concentrate urea while minimizing water loss. 3 rats were imaged with both agents; HMCP served as a control since this molecule is not expected to be affected by urea transporters. The methods for inducing diuresis and antidiuresis were identical to those described previously \cite{morze_ut}. To induce antidiuresis, the rats were deprived of food and water for an overnight period of $16$ hours. To induce osmotic diuresis, the rats were first deprived of food and water for $16$ hours and then allowed free access to aqueous glucose (10\% by mass) solution for 9 hours. In each experiment, the urea injection (3 mL, $150$ mM hyperpolarized [\textsuperscript{13}C,\textsuperscript{15}N$_2$]urea) was performed at least 2 hours prior to the HMCP injection (3 mL, $125$ mM hyperpolarized HMCP). The $T_2$ mapping sequence was initiated 40 s after the beginning of injection with identical acquisition parameters aside from a 4.5 kHz resonance frequency offset between urea and HMCP.

 \paragraph{Statistical Analysis}
Both kidneys were manually delineated on all \textsuperscript{13}C images after alignment and denoising, and only pixels within the kidney were included in analysis.  Following pixel selection, a $T_2$ distribution was computed for each pixel. A single mean $\langle T_2 \rangle$ value was calculated at each pixel from this distribution by calculating the first moment over the full distribution ($\alpha = 0.3$ s, $\beta = 20$ s). The use of the full distribution was motivated by the difficulty in comparing two molecules with differing relaxation properties. 4 maps were computed for each of the 3 animals: $\langle T_2 \rangle$ maps for urea and HMCP in diuresis and antidiuresis states. Histograms including all pixels binned by $\langle T_2 \rangle$ were plotted for each agent to visualize the effect of diuresis and antidiuresis states. For quantitative comparison, the mean of the upper-90\textsuperscript{th} percentile was computed for each animal. Due to the inward $\langle T_2 \rangle$ gradient observed with both agents, this operation selected pixels within the center of the kidney. Two paired t-tests were then performed: the first compared the mean upper-90\textsuperscript{th} percentile of $\langle T_2 \rangle$ of urea between diuresis and antidiuresis, and the second compared the mean upper-90\textsuperscript{th} percentile of $\langle T_2 \rangle$ of HMCP between diuresis and antidiuresis. $p<0.05$ was used as the significance criteria.

\paragraph{$^1$H MRI}
Animals were imaged in the coronal and axial planes using a $T_2$-weighted $^1$H fast spin echo (FSE) sequence with $TR=1$s, $TE=$100 ms, 32 echoes, 0.8 mm in-plane resolution.

\subsection*{High Resolution \textsuperscript{13}C Imaging Techniques}

\paragraph*{Multiple Echo Time Averaging}

The persistence of signal across multiple echo times allowed for image averaging to increase the effective $SNR$. This summation was performed after motion correction and $SVD$ denoising. 

\paragraph*{3D Imaging}
As an alternative multi-exponential fitting, the MRI acquisition may be designed with a long echo time to filter out the short-$T_2$ signal components. One method of achieving this is using 2 spatial dimensions  of phase encoding (for the acquisition of a full 3D image) with a rasterized Cartesian ordering in which the echo time is half the total acquisition time.  A 3D SSFP acquisition was initiated at 20 s, 25 s, and 30 s after the beginning of urea injection. This sequence used a large flip angle ($\theta=120^\circ$) and extremely long echo time (3.5 s) allowing for editing out the vascular signal with enhancement of the long-$T_2$ filtrate. The un-aliased $FOV$ was chosen to cover only the kidneys in the 2 phase encoded dimensions. Since the images were sampled below the Nyquist cutoff frequency in these dimensions, the signal filtering of the ultra long echo time acquisition was utilized to reduce aliasing from the vascular signal.  A $(42,42,14)$ acquisition matrix was acquired over a $(5,5,1.7)$ cm $FOV$ in (L-R,S-I,A-P) coordinates yielding a 1.2 mm isotropic pixel length. 588 phase encodes were acquired with $TR=12$ ms for a 7 s total scan time.

 \section*{Results}
 \subsection*{\textsuperscript{13}C Urea / BSA-GdDTPA Relaxometry }
The $T_1$ relaxivity of the BSA-GdDTPA complex on \textsuperscript{13}C urea was estimated to be $77\pm10$ mM$^{-1}$s$^{-1}$ with respect to the BSA carrier, or $3.1\pm.4$ mM$^{-1}$s$^{-1}$ per GdDTPA chelate ($R^2=.97$) from saturation recovery experiments. Fig. 3e shows the measured $T_1$ relaxivity curve with error bars indicating $T_1$ measurement uncertainty from the intrinsic signal to noise ratio. The estimated rat blood volume was $27\pm3$ mL \cite{lee_rat_bloodvol}. Therefore, at the expected \textit{in vivo} [BSA-GdDTPA] of $\sim0.024$ mM, the urea $T_1$ should be approximately 0.5 s. Although this calculation is extremely rough since it ignores circulation or potentially differing relaxivity \textit{in vivo} and \textit{in vitro}, the 7 s delay should have been adequate for any hyperpolarized \textsuperscript{13}C urea to undergo several $T_1$ time constants of decay leading to large polarization loss when in contact with BSA-GdDTPA.

\begin{figure}[!h]
	\centering
	\includegraphics[width=3.5in]{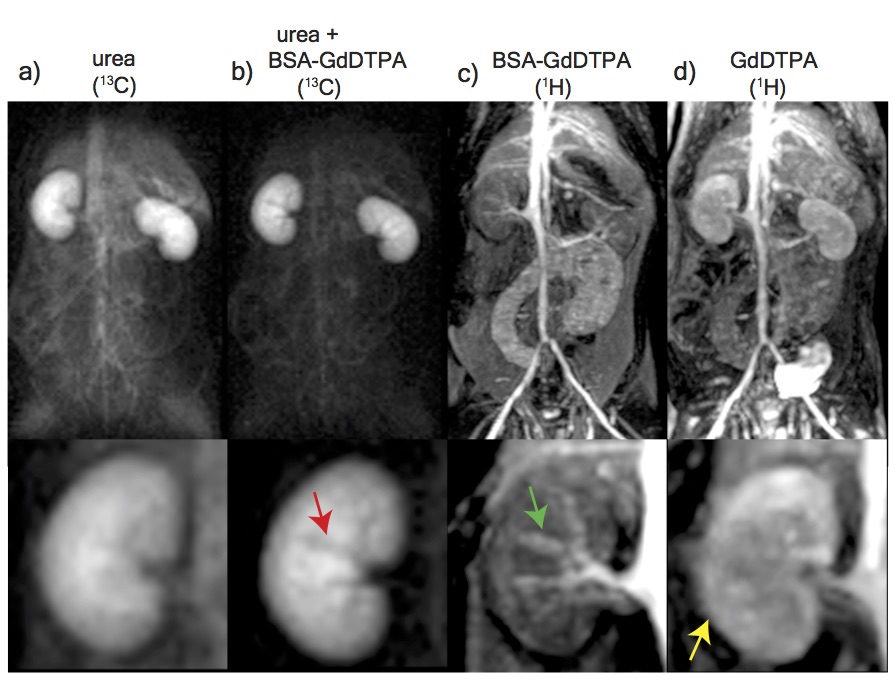}
\caption{{\bf Differential attenuation of the \textsuperscript{13}C urea signal after a chase injection of the intravascular agent BSA-GdDTPA. } Large field-of-view images are shown on the top, and the bottom panels are zoomed to the kidney. a) First time-point \textsuperscript{13}C urea MRI image. b) \textsuperscript{13}C urea MRI acquired 8 s after BSA-GdDTPA infusion shows strong suppression of the vascular signal and interlobular arteries (red arrow). c) \textsuperscript{1}H MRI acquired 5 minutes after BSA-GdDTPA infusion. The interlobular arteries show positive contrast in this image (green arrow). d) \textsuperscript{1}H MRI acquired 5 minutes after GdDTPA infusion. In contrast to BSA-GdDTPA (mass $\sim85$ kDa), GdDTA (mass $\sim938$ Da) is freely filtered at the glomerulus (yellow arrow). 
}
\end{figure}

Figure 2 shows a first echo \textsuperscript{13}C urea image with (Fig. 2a) and without (Fig. 2b) the chaser injection of the macromolecular BSA-GdDTPA relaxation agent. This image shows a large suppression of the vascular \textsuperscript{13}C urea signal throughout as well as darkening of the interlobular branches of the renal artery (Fig 2b). The darkening of the renal arterial branches indicates that the 7 s delay was adequate not only for both bolus arrival of the BSA-GdDTPA to the kidneys, but also for the BSA-GdDTPA to cause substantial \textsuperscript{13}C urea polarization loss. These same arterial branches show up bright in a $T_1$-weighted $^1$H image (Fig 2c). No renal perfusion was detected with the BSA-GdDTPA agent up to 5 minutes after initial infusion. In contrast, the post-GdDTPA image (Fig 2d) shows renal perfusion and bladder accumulation since this agent is freely filtered at the glomerulus \cite{zhang_mri_nephro}.

\begin{figure}[!h]
	\centering
	\includegraphics[width=3.7in]{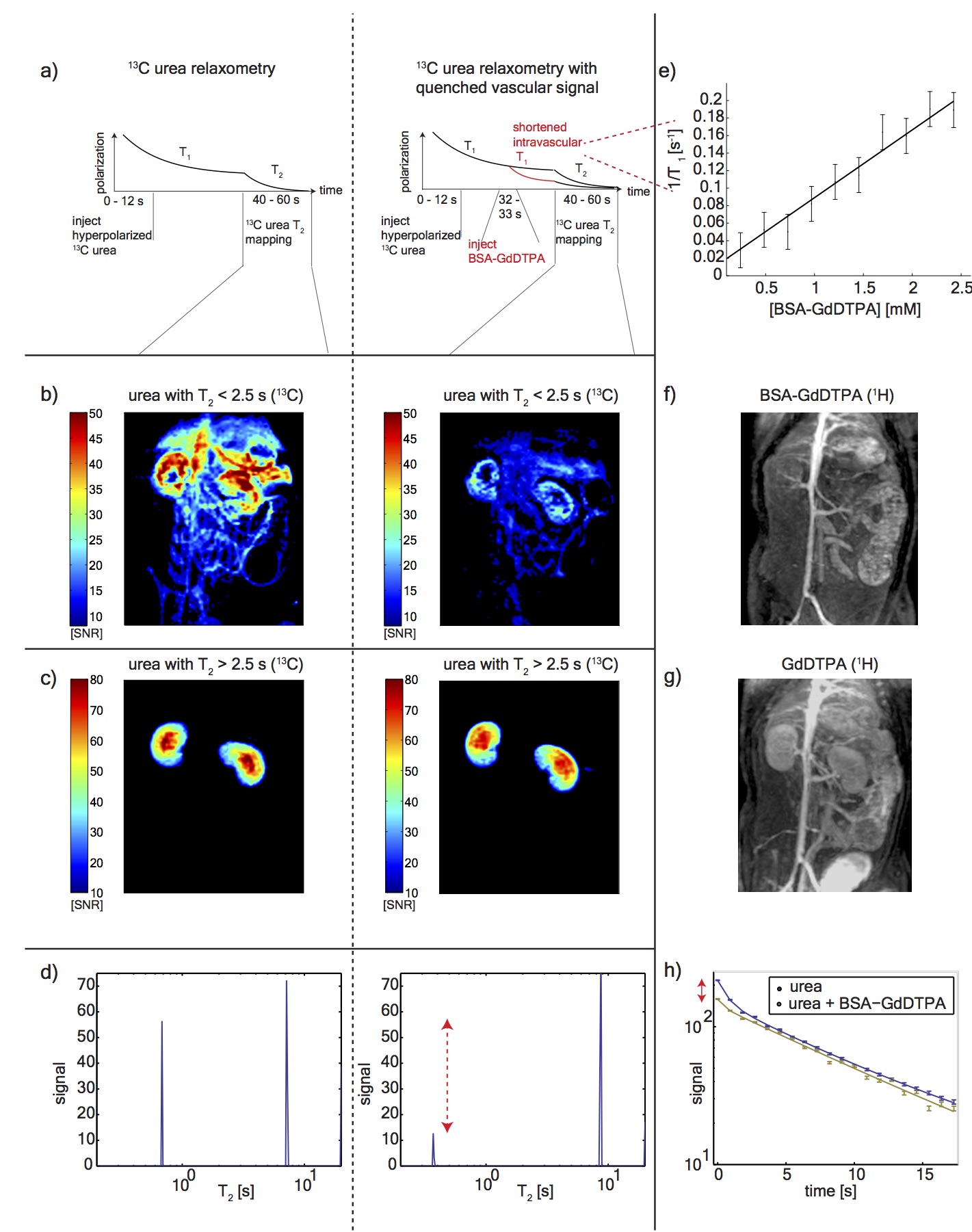}
\caption{{\bf \textsuperscript{13}C urea $T_2$ relaxometry after quenching the vascular signal.}
 a) Timeline of the substrate injections and imaging. The left column shows images from the control experiment, and the center column shows the post BSA-GdDTPA images. b) \textsuperscript{13}C urea signal outside of the kidneys has $T_2$ less than 2.5 s which was strongly attenuated by BSA-GdDTPA. c) The long-$T_2$ urea signal component was confined to the kidneys and was unaffected by the BSA-GdDTPA chaser. d) $T_2$ distributions of single pixels selected from the center of the kidneys showing this short-$T_2$ signal attenuation (red arrow). e) The \textsuperscript{13}C urea relaxation rate $1/T_1$ decreases linearly with BSA-GdDTPA concentration with a slope of $77\pm10$ mM$^{-1}$s$^{-1}$. f) $T_1$ weighted \textsuperscript{1}H imaging 5 minutes post BSA-GdDTPA infusion. g) $T_1$ weighted \textsuperscript{1}H imaging post GdDTPA infusion. h) Single pixel $T_2$ decay curves (corresponding to the distributions in d). }
\end{figure}


The dynamic \textsuperscript{13}C urea images acquired under $T_2$ decay conditions initially showed greatly reduced signal at early echo times when accompanied by the BSA-GdDTPA chaser. At later echo times, however, the images converge and look nearly identical. This effect is most easily visualized in the supplemental video S1: by the 7\textsuperscript{th} time point (corresponding to a 6.8 s echo time) only urea within the kidneys is visible in both images, and the signal's spatial variation is nearly identical in the images with and without the BSA-GdDTPA chaser. The similarity persisted in all experiments until the end of imaging acquisition. At this echo time, both images show urea signal throughout the cortex and medulla, and a dark-rim is present at the outer stripe of the outer medulla. The later echo images were acquired up to 25 s after the infusion of the BSA-GdDTPA chaser and provide strong evidence that the slowly decaying \textsuperscript{13}C urea signal component emanated from regions inaccessible to the BSA-GdDTPA.

\begin{figure}[!h]
	\centering
	\includegraphics[width=3.7in]{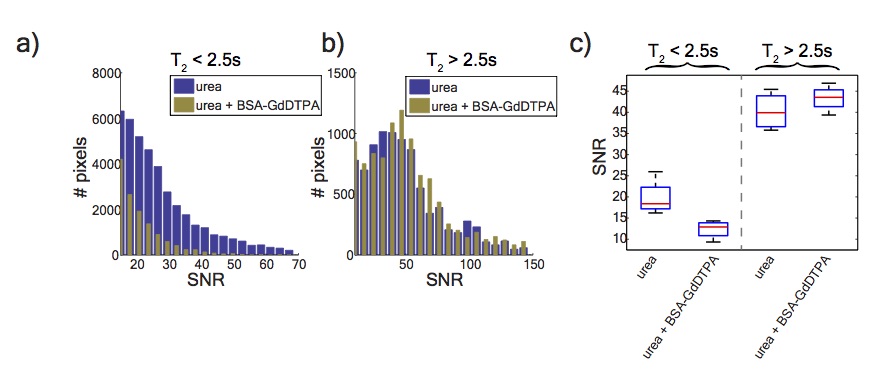}
\caption{{\bf \textsuperscript{13}C urea $T_2$ relaxometry after quenching the vascular signal.}
  Pixel distributions of the short-$T_2$ (a) and long-$T_2$ (b) of the \textsuperscript{13}C urea signal with and without BSA-GdDTPA chaser. All pixels with first time point SNR $>10$ from 4 animals are displayed in these histograms. c) shows the range of the mean of these signal components for the 4 animals scanned. A paired t-test show significant ($p<0.05$) attenuation of the short-$T_2$ by the BSA-GdDTPA chaser (shown in the left 2 boxes).  }
\end{figure}

Multi exponential relaxometry quantified this dissimilarity at early echoes and similarity at late echoes. Fig 3b and 3c show the short-$T_2$ and long-$T_2$ signal components, respectively, with the latter appearing identical in the \textsuperscript{13}C urea images with and without BSA-GdDTPA. Fig 3h shows a single decay curve from a pixel selected from the center of the kidney with semi-log axes. The two decay modes show up as peaks in the $T_2$ distributions in Fig 3d, with the short-$T_2$ component being significantly reduced by the BSA-GdDTPA chaser. Binning the all the pixels from all animals scanned also clearly showed this differential attenuation of the short- and long-$T_2$ signal components (Fig. 4a, 4b). The mean of the short- and long-$T_2$ components for the 4 animals is shown in the box plot in Fig. 4c. The whiskers represent the minimum and maximum of the mean of the short- and long-$T_2$ maps. The paired t-test showed significant ($p<.05$) diminishing of the SNR of the short-$T_2$ \textsuperscript{13}C urea signal component over all animals.

\subsection*{Diuresis / Antidiuresis Relaxometry}

Fig. 5 shows the long-$T_2$, extravascular \textsuperscript{13}C urea. During osmotic diuresis, the urea remains largely in the outer medulla and cortex at the imaging start time used (40 s after the beginning of a 12 s injection). During antidiuresus, a larger fraction of the urea is collected in the inner medulla and papilla consistent with the enhanced inner-medullary urea pool during antidiuresis indicative if the UT-A1 and UT-A3 transporters (Fig 5, left). When the acquisition delay allowed for inner-medullary urea accumulation, large $T_2$ increases were observed due to a strong inward $T_2$ gradient were observed with the long-$T_2$ component of \textsuperscript{13}C urea (for example, see Fig 10 for a $T_2$ distribution of the extravascular \textsuperscript{13}C urea). Given this $T_2$ gradient, relaxometry is a sensitive detection method for observing urea transporter facilitated concentrating ability. Supplementary video S2 shows the dynamic \textsuperscript{13}C urea at multiple echo times. During antidiuresis, not only is a greater inner medullary accumulation of the \textsuperscript{13}C urea observed, but the signal persists to very late echo times.

\begin{figure}[!h]
	\centering
	\includegraphics[width=3.2in]{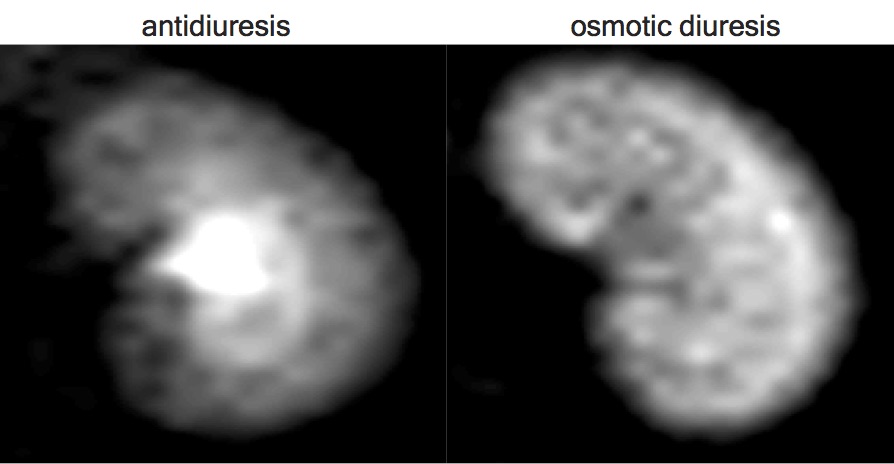}
\caption{{\bf Extravascular \textsuperscript{13}C urea during antidiuresis and diuresis.}
  A larger fraction of the urea is collected in the inner medulla and papilla consistent with the enhanced inner-medullary urea pool during antidiuresis indicative if the UT-A1 and UT-A3 transporters (left). During diuresis, the urea is primarily in the cortex and outer medulla (right) at the time of imaging. }
\end{figure}

\begin{figure}[!h]
	\centering
	\includegraphics[width=3.7in]{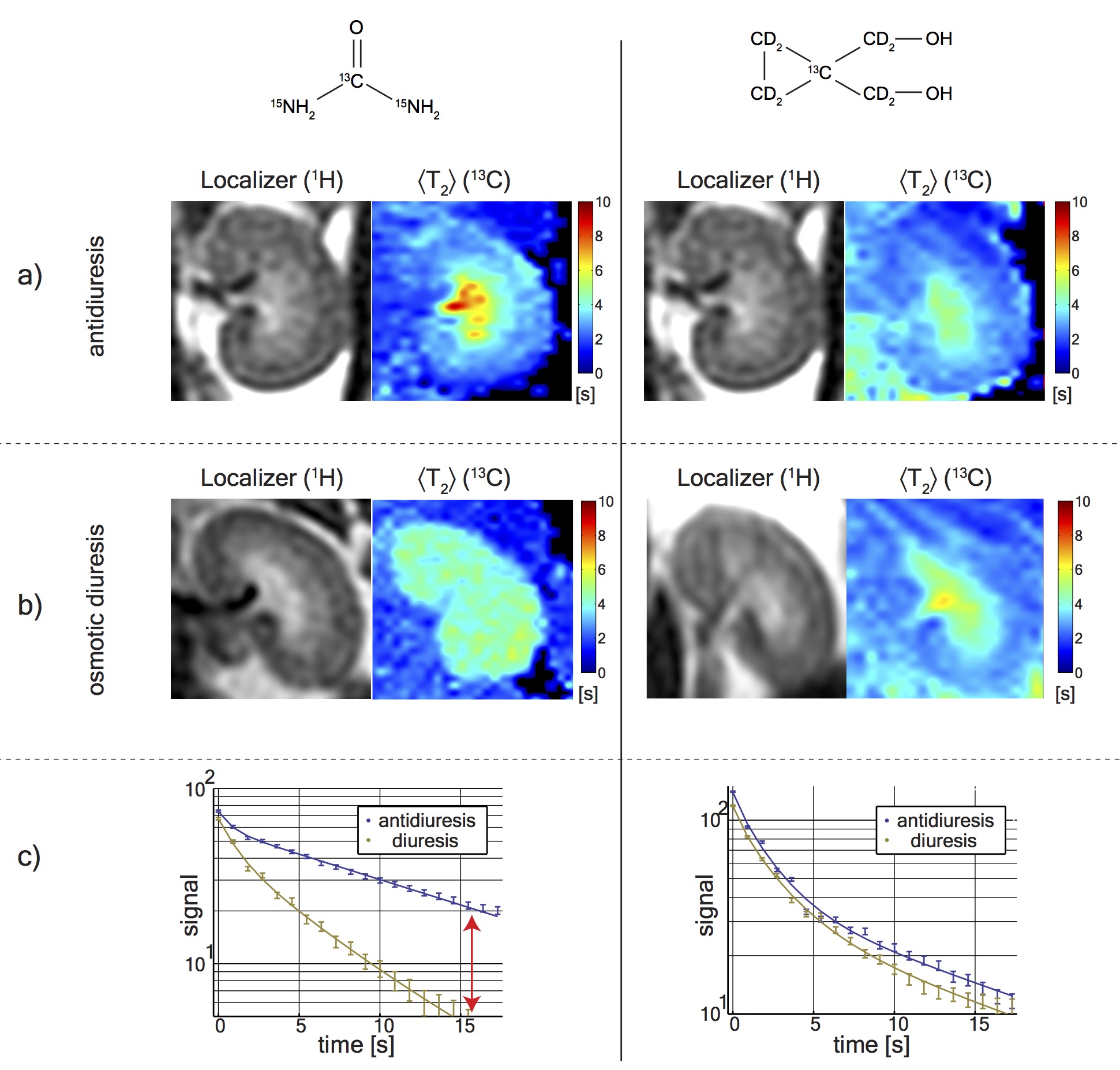}
\caption{{\bf Hyperpolarized \textsuperscript{13}C urea and HMCP relaxometry during antidiuresis and diuresis.}
 Hyperpolarized \textsuperscript{13}C urea and HMCP images are shown the left and right columns, respectively. HMCP showed increased medullary $T_2$ but the effect did not vary between diuresis (b) and antidiuresis (a) states. c) decay curves selected from a single pixel in the inner medulla. The red arrow shows the persistence of signal to late echo times. }
\end{figure}

Fig. 6 shows $T_2$ exponential relaxometry performed with both imaging agents in rats on induced diuresis and antidiuresis. Rather than selecting the short- or long-$T_2$ components, the mean $\langle T_2\rangle$ (calculated by equation \ref{eq_meant2} with $\alpha = 0.3$ s, $\beta = 20$ s) was used to simplify analysis. A large $\langle T_2\rangle$ increase was observed in the inner medulla and renal papilla with \textsuperscript{13}C urea during antidiuresis (Fig. 6a). This $\langle T_2\rangle$ increase in the kidneys' central region was not observed in diuresis (Fig. 6b). The signal distribution and $\langle T_2\rangle$ of HMCP did not change significantly between antidiuresis and diuresis (Fig. 6a,b, right). This effect can also be seen in the pixel histogram $\langle T_2\rangle$ distributions (selected from pixels within the kidney) in Figure 7.  

Both agents showed $\langle T_2\rangle$ lengthening in the inner medulla and papilla. Unlike \textsuperscript{13}C urea, HMCP showed a small $\langle T_2\rangle$ reduction in the cortex and outer medulla compared to the blood. HMCP had a much higher $\langle T_2\rangle$ in the vascular pool (4 s) compared to urea (1 s). This effect is most easily visualized in the outer margins of the images in Fig. 6a,b (right). With \textsuperscript{13}C urea, $\langle T_2\rangle$ values greater than 2 s were only observed within the kidney.


\begin{figure}[!h]
	\centering
	\includegraphics[width=3.7in]{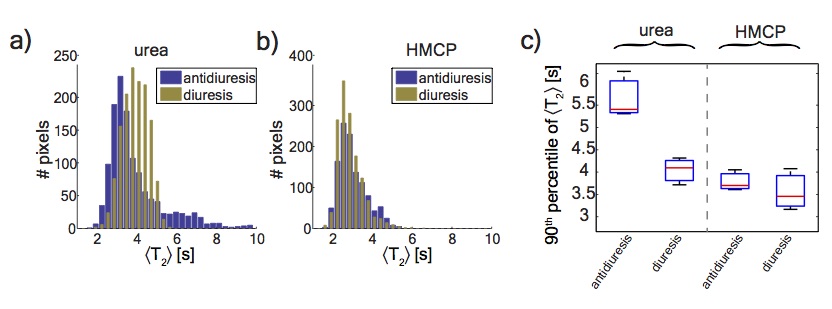}
\caption{{\bf Pixel $\langle T_2\rangle$ distributions of \textsuperscript{13}C urea and HMCP relaxometry during diuresis and antidiuresis.}
a) \textsuperscript{13}C urea and b) HMCP pixels from the kidneys of 3 animals. The distribution extends to high $\langle T_2\rangle$ values for \textsuperscript{13}C urea in antidiuresis indicating \textsuperscript{13}C urea concentration in the inner medullary collecting ducts and medullary interstitium. This effect was not observed with HMCP. c) the mean upper 90\textsuperscript{th} percentile plotted for all animals.    
   }
\end{figure}

\subsection*{High Resolution \textsuperscript{13}C Imaging Techniques}

After image alignment, denoising, and echo time averaging, signal in the ureters could be observed in osmotic diuresis conditions with both \textsuperscript{13}C urea and HMCP (see Fig 8). The inner lumen of the rat ureter has a diameter between 50 and 150 $\mu$m \cite{hicks_ureter}, so a 1 mm width pixel (acquired perpendicular plane to the ureter axis) contains on the order of 10 nL intra-ureter fluid. Although the signal was extremely faint, the persistence of signal allowed for detectable ureter structure. HMCP had a much stronger signal than \textsuperscript{13}C urea in the ureters likely due to its longer $T_1$. Ureters were only visible in 2 of the 3 diuresis scans for HMCP and 1 of 3 scans for \textsuperscript{13}C urea. The long $T_1$ of HMCP enabled urinary bladder collection to be imaged at over 3 minutes post injection (image not shown).

\begin{figure}[!h]
	\centering
	\includegraphics[width=3.2in]{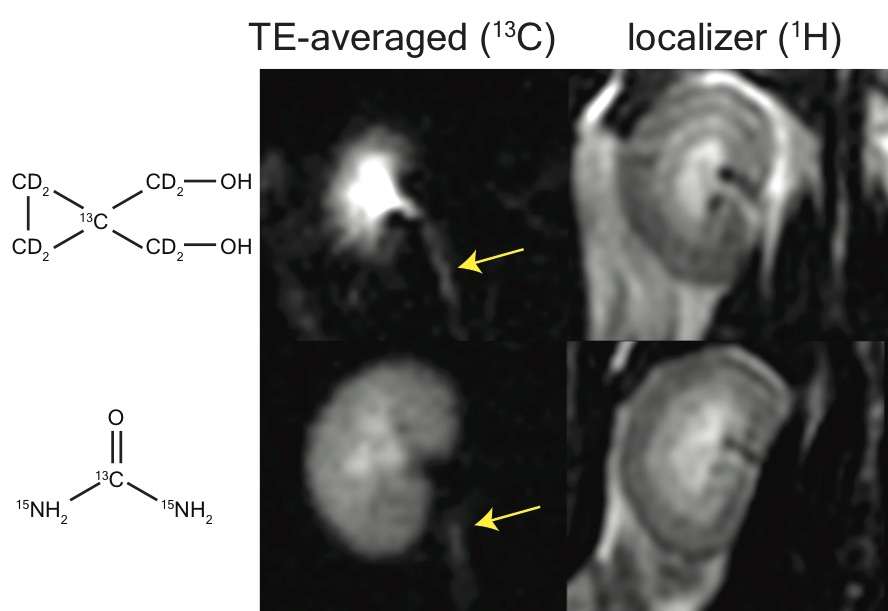}
\caption{{\bf \textsuperscript{13}C ureter imaging during diuresis.}
10 echo times (from 5 s to 14 s) were averaged after alignment and denoising to improve signal. Ureters could be observed in diuresis with both urea and HMCP (arrows). High urea reabsorption and reduced UT-A1 and UT-A3 activity leads to substantial cortical and outer medullary signal compared to HMCP. 
}
\end{figure}

Fig. 9 shows 3D images acquired at 1.2 mm isotropic resolution (1.73 mm$^3$ pixel volume). These images are zoomed to a single kidney. The efficacy of the blood pool suppression is evidenced by the low background signal as well as the dark interlobular arteries (magenta arrows on the image panels). The outer stripe of the outer medulla (OSOM) enhanced later than the cortex and inner stripe of the outer medulla (ISOM, see the 20 s and 25 s time points). Once inner medullary accumulation occurred, the inner medulla (IM) and papilla were bright due to the sequence weighting (see simulations in Fig 9a, right). These images highlight the sensitivity in rats of imaging timing for desired contrast agent distribution. These images also represent, to the authors' best knowledge, the first \textit{in vivo} hyperpolarized \textsuperscript{13}C images acquired at sub-2 mm$^3$ isotropic resolution.     



\begin{figure}[!h]
	\centering
	\includegraphics[width=3.7in]{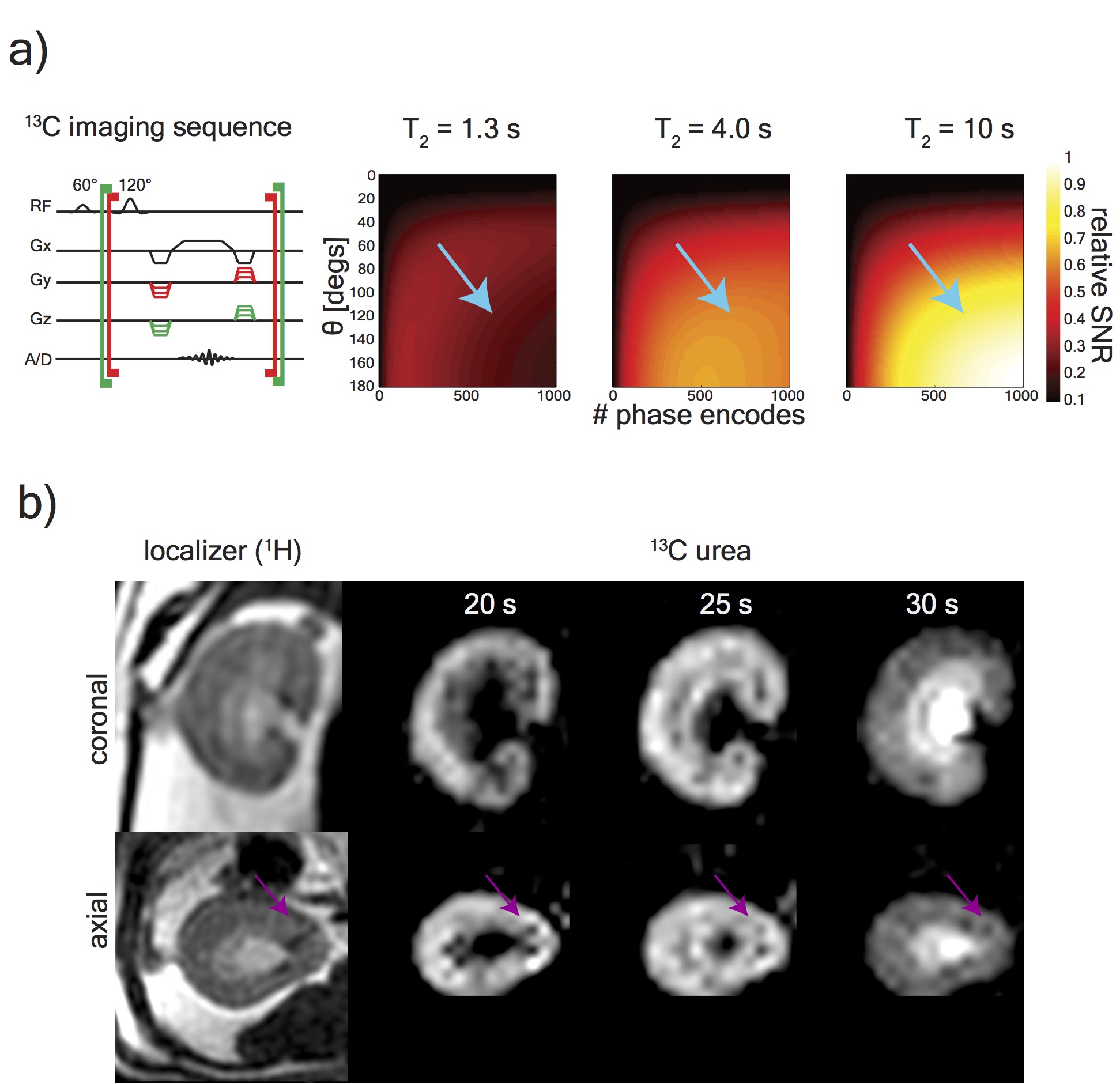}
	\caption{{\bf \textsuperscript{13}C urea imaging at sub-2 mm$^3$ isotropic resolution}
a) Sequence design (left) and signal simulations (right 3 panels) of a 1.7 mm$^3$ isotropic resolution image. Simulations show the signal response expected from regions with $T_2=1.3$ s (vascular pool), $T_2=4$ s (cortex / outer medulla), and $T_2=10$ s (inner medulla / papilla). Blue arrows indicate sequence parameters used, and thus the signal response expected in the images. b)  The long effective echo time (4 s) of a 3D acquisition was utilized for blood pool signal suppression and for encoding the \textsuperscript{13}C urea at 1.2 mm isotropic resolution. b) \textsuperscript{13}C urea images acquired at three different delay times after three different injections. Blood pool suppression is evidenced by the dark interlobular arteries visible on both the \textsuperscript{13}C urea and \textsuperscript{1}H fast spin echo images (magenta arrows). }

\end{figure}

\section*{Discussion}

\paragraph{Comparison of \textsuperscript{1}H and \textsuperscript{13}C MRI of the Kidney}

Hyperpolarized \textsuperscript{13}C MRI is a background-free imaging method that allows direct detection the labeled contrast agent in the kidney, whereas Gd-enhanced \textsuperscript{1}H MRI utilizes the enhanced relaxation rate of water for indirect detection. The latter method, therefore, always has signal contribution from regions without contrast agent, but this contribution is typically minimized with strong $T_1$ weighting of the acquisition. The experiments in this study highlight that relaxation is also an extremely important factor for the analysis of hyperpolarized \textsuperscript{13}C images. Regions of hypo-intense signal may stem not only from low contrast agent concentration, but may also arise from $T_1$ or $T_2$ shortening, and this effective contrast can be exacerbated by the acquisition. 

In both hyperpolarized \textsuperscript{13}C and Gd-enhanced \textsuperscript{1}H MRI, the molecular label carrier influences the renal contrast agent distribution. Since GdDTPA is not significantly reabsorbed along the nephron \cite{choyke_kidney,lee_multicompartment}, the signal likely arises primarily from the lumen of the tubules and collecting ducts as well as the glomerular blood supply. A significant fraction of urea in the filtrate, however, is reabsorbed along the nephron. Fig. 8 shows qualitatively the high reabsorption of urea compared to HMCP in rats on induced diuresis. Urea (mass $\sim63$ Da) and HMCP (mass $\sim110$ Da) are both freely filtered at the glomerulus, but the late echo time urea image (Fig 8, bottom left) still shows substantial cortical signal. It is worth noting that other commonly used endogenous molecular probes for hyperpolarized \textsuperscript{13}C MRI such as pyruvate and lactate are also freely filtered but are nearly completely reabsorbed under normal conditions \cite{hohmann_lactate_kidney, anderson_pyruvate_diabetes}, so significant collection in the inner medulla and renal pelvis would not be expected. Urea is an endogenous metabolic waste product, and HMCP is a non endogenous molecule, so the kidney attempts to excrete both of these agents.

\begin{figure}[!h]
	\centering
	\includegraphics[width=3.5in]{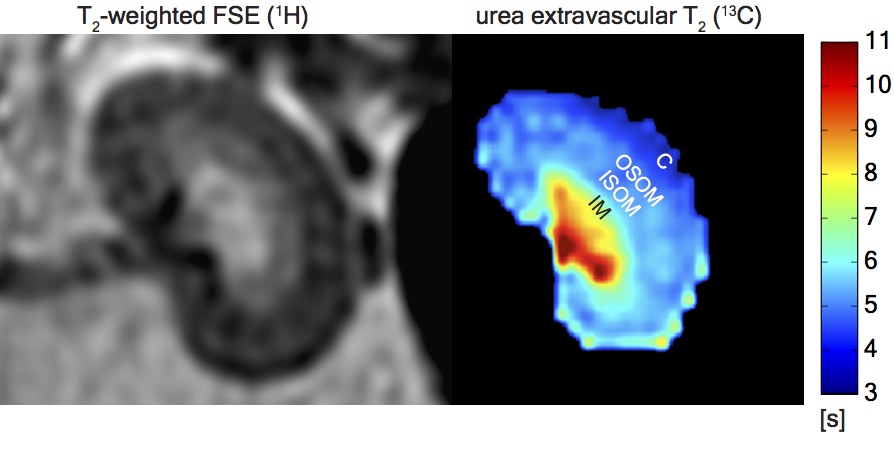}
	\caption{{\bf $T_2$ of extravascular \textsuperscript{13}C urea}
$T_2$-weighted \textsuperscript{1}H localizer is shown on the left. The right image is a $T_2$ map isolated to the renal filtrate ($T_2>2.5$ s). The cortex (C) and inner stripe of the outer medulla (ISOM) gave lower $T_2$ values than the outer stripe of the outer medulla (OSOM) and inner medulla (IM). These $T_2$ variations are potentially indicative of slow \textsuperscript{13}C urea exchange between the medullary interstitial fluid and the vasa recta.   
}
\end{figure}

The spatial variation of \textsuperscript{13}C $\langle T_2 \rangle$ is compared with a $T_2$-weighted spin echo \textsuperscript{1}H MRI in Fig. 10. In this figure, the mean $\langle T_2 \rangle$ was calculated to represent that of the extravascular urea pool ($T_2>2.5$ s). The \textsuperscript{1}H $T_2$ correlates well with water content \cite{kundel_kidneyt1} and increases with edema \cite{hueper_renalt2}, while \textsuperscript{13}C urea undergoes a strong relaxation enhancement from erythrocytes \cite{reed_n15urea}. Unlike contrast-enhanced imaging with GdDTPA which is not significantly reabsorbed, the extravascular urea signal likely arises from the tubular lumen as well as the interstitial fluid (IF) due to high permeability along the proximal tubule, the thin limbs of the loop of Henle, and the inner-medullary collecting ducts \cite{lassiter_micropunc,marsh_urea_review}. The relative contribution of tubular and interstitial pools on the short timescale of the \textsuperscript{13}C imaging experiment is unknown, but the spatial $\langle T_2 \rangle$ variation (Fig. 10, right) could suggest a large contribution from the IF. The cortex (C) and inner stripe of the inner medulla (ISOM) showed $\langle T_2 \rangle$ values less than the outer stripe of the outer medulla (OSOM) and inner medulla (IM). A potential explanation is the urea exchange between the medullary interstitium and the peritubular microvascular supply. Urea within the IF is in exchange with the vasa recta, a process which is critical for efficiency of the countercurrent exchange mechanism. The density of the peritubular vascular network differs within the various renal segments. The peritubular vascular tissue fraction in the cortex and ISOM is nearly double that of the OSOM and IM \cite{kriz_medullary_circ,garcia_microvasc_kidney}. The medullary microvascular blood supply is sparse, receives less than 1\% of the total renal blood flow \cite{guyton_book}, and occupies less than 0.4\% of the total tissue volume \cite{garcia_microvasc_kidney}. However, the vasa recta are highly permeable to urea, and the apparent $\langle T_2 \rangle$ lengthening of the \textsuperscript{13}C urea isolated to the filtrate in the sparse miscrovascular regions (OSOM, IM) and shortening in the dense microvascular regions (C, ISOM) potentially indicate that the $T_2$ variations observed indicate a slow exchange of urea between the interstitium and the microvascular supply.

\paragraph{Comparison with prior hyperpolarized \textsuperscript{13}C NMR measurements}

The \textsuperscript{13}C urea $T_1$ relaxivity of the macromolecular relaxation BSA-GdDTPA measured here ($3.1\pm.4$ mM$^{-1}$s$^{-1}$ per GdDTPA, $B_0 = 3$ T) is more than an order of magnitude higher than that reported with Gadodiamide and [1-\textsuperscript{13}C]pyruvate ($0.19\pm.01$ mM$^{-1}$s$^{-1}$, $B_0 = 4.7$ T) \cite{smith_gdchaser}. This is likely attributable to the increased relaxivity from the slow correlation time of the macromolecule Gd carrier \cite{werner_review} but may also signify some preferential binding of urea to albumin.

Based on measurements in this and prior studies, \textsuperscript{13}C urea (with \textsuperscript{15}N labeling) experiences strong relaxation enhancement in plasma ($T_2 = 11$ s) and whole blood ($T_2 = 5$ s) compared to aqueous solution ($T_2 = 24$ s) \cite{reed_n15urea}. Although the strongest \textit{in vitro} relaxation enhancement was observed in whole blood, this value is more than double that measured \textit{in vivo} ($T_2 = 1.2$ s) potentially indicating a large decrease in the apparent $T_2$ due to flow. Relaxation enhancement in erythrocytes is likely due to a combination of high blood viscosity and high erythrocyte membrane permeability \cite{brahm_urea_rbc}; the latter is mediated by the erythrocyte isoform of the urea transporter UT-B. A 20\% reduction in the diffusion coefficient of urea has been reported in UT-B expressed tissue xenografts compared to controls likely indicating an increased intracellular pool size \cite{patrick_urea_transgene}. Controlled experiments in erythrocytes reported a changes in the resonant frequency \cite{potts_urea_rbc_nmr} and reduction of the longitudinal $T_1$ relaxation time \cite{pages_urea} of the intracellular \textsuperscript{13}C urea. These results are concordant with a shortening of the vascular $T_2$ due to erythrocyte permeability.

The vascular $T_2$ of HMCP measured here ($T_2=4$ s at $B_0=3$ T) is in agreement with a "worst case" value ($T_2=1.3$ s at $B_0=2.35$ T) reported previously using SSFP \cite{SvenssonC13SSFP} but is over an order of magnitude higher than \textit{in vivo} measurements ($T_2=0.4$ s at $B_0=9.4$ T) reported using adiabatic refocusing pulses with surface coils \cite{kettunen_spinecho} suggesting further study in the dependence of apparent $T_2$ measurements on $B_0$, contrast agent uptake in tissue, and acquisition type.


\paragraph{Experimental Limitations}

Multicomponent $T_2$ relaxometry has well-known difficulties in accurately resolving closely-spaced $T_2$ values at the SNR, scan times, and echo spacing permissible on clinical MRI scanners \cite{graham_t2_criteria}. In this study, the SNR limitation was somewhat exacerbated by the polarization loss from $T_1$ decay during the long delay periods between injection and imaging necessary for renal contrast agent accumulation. The observed $T_2$ decay times were significantly longer than those typical of \textit{in vivo} $^1$H MRI which permitted the coarse temporal sampling required by the single-shot, 1 mm resolution planar readout. Furthermore, $T_2$ differences between intra- and extravascular \textsuperscript{13}C urea differed by a factor of at least 4. Decreasing the resolution would allow finer temporal sampling and increased SNR for the stabilization of $T_2$ distribution estimation \cite{graham_t2_criteria}. $T_2$ values less than the sampling time of 910 ms are not expected to be resolved accurately. In non-selective \textit{in vivo} CPMG experiments performed after infusion of hyperpolarized \textsuperscript{13}C urea with finer temporal sampling (10 ms), we observed approximately 30\% of the total signal had a $T_2$ of 300 ms or less.

In each experiment, the rat was given two 3 mL injections of hyperpolarized contrast agents spaced at least 2 hours apart. For the macromolecular relaxation experiments, the rat was given 2 additional 1 mL injections of Gd contrast. The total injected volume equals approximately 1/4 the animals' total blood volume and had an unknown affect on cardiac output, glomerular filtration rate (GFR), and renal concentrating capacity. Additionally, the urea mixture contained glycerol which, in large doses, is known to cause GFR reduction in rats \cite{lin_glycerl_gfr}. Although it is difficult to conclude that this large aggregate injection volume had no transient effect on renal function, we did not observe imaging evidence of differing renal concentrating capacity with repeat injections when the imaging was started at a constant duration after the injection.

Polarization variability will lead to random errors in comparing absolute SNR between experiments. Although prior measurements showed less than a 15\% variability in \textsuperscript{13}C urea polarizations 
when polarization and transport time are kept consistent \cite{reed_n15urea}, this potential random error will almost certainly be minimized with the use of automated transport injectors \cite{cheng_powerinjector} and magnetically shielded transport pathways \cite{milani_magnetictunnel}.

Systematic errors may arise in quantitative $T_2$ mapping due to transmitter strength mis-calibrations when using the transient phase of the SSFP signal. As derived by Scheffler, the exponential decay envelope of the signal is described by the positive eigenvalue \cite{scheffler_ssfp}
\begin{equation}
\lambda_1 = \frac{1}{2}\left( \left(E_1-E_2\right)\cos\theta +\sqrt{4E_1E_2 + \left(E_1-E_2\right)^2\cos^2\theta} \right), 
\end{equation}
with $E_1 = e^{-TR/T1}$, $E_2 = e^{TR/T_2}$. The transmitter offset may be modeled as $\theta = 180^\circ + \delta \theta$, and a Taylor expansion of $\lambda_1$ gives 
\begin{equation}
\lambda_1\approx a_0 + a_1 \delta\theta + a_2 \delta\theta^2 + ...
\end{equation}  
with 
\begin{align}
a_0 &= E_2 \\
a_1 &= 0 \\
a_2 &= \frac{E_2\left(E_1-E_2\right)}{E_1+E_2}
\end{align}
Non-ideal $\pi$ pulses will cause some apparent lengthening of the measured $T_2$ by introducing some $T_1$ weighting. However, these errors show up only as second or higher even order terms of $\delta \theta$. The second order term is minimized when $E_1\approx E_2$, and this condition is expected to be better approximated in the longer $T_2$ regions. Assuming $T_2 =1.5$ s, $T_1 = 20$ s, flip angle errors $\delta\theta/\theta$ up to $20\%$ cause the apparent decay time to differ from $T_2$ by less than $10\%$.

\paragraph{Potential Clinical Utility}

Although $T_2$-weighted imaging has been a standard clinical MRI evaluation for three decades \cite{hennig_rare}, only a few studies have investigated $T_2$ contrast for \textit{in vivo} hyperpolarized \textsuperscript{13}C agents imaging \cite{yen_t2, reed_n15urea,kettunen_diffusion}. In addition to chemical shift \cite{GolmanMetabolic,witney_kinetics,schulte_satrecover,josan_dse,lau_multislice,ramirez_radial} and diffusion \cite{koelsch_reed_diff,patrick_urea_transgene,larson_supersteam,kettunen_diffusion,schilling_diffusion} sensitive imaging techniques, $T_2$ relaxometry could be a very useful tool for probing the microenvironment of hyperpolarized \textsuperscript{13}C molecules \textit{in vivo} since it yields high signal thus allowing for high resolution encoding.

Given the importance of urea in the urine concentrating mechanism, high resolution imaging of renal urea handling could be a powerful tool for the investigation of renal physiology. This imaging technique could be applicable to the monitoring diuretic drugs which act on urea transporters \cite{esteva_uta,yao_utb,sands_uta1_diuretics} or to study the effects of antineoplastic drugs whose side effects include reduced urea concentrating ability \cite{safirstein_cisplatin}. Radiologically, this method could address the inherent difficulty of renal perfusion evaluation on patients with impaired renal function and chronic kidney disease since virtually all commonly used iodinated CT contrast agents and gadolinium-based MRI contrast agents pose some hazard of acute renal failure in these patients \cite{parfrey_iodinated_contrast,morcos_kidney}. Although urea clearance is well known to be an inaccurate marker for GFR estimation due to its significant reabsorption \cite{marsh_urea_review}, \textsuperscript{13}C urea MRI could provide a qualitative assessment of renal perfusion such as is regularly performed  on transplantation candidates, before and after ablation for renal cell carcinoma, and for the assessment of congenital urological abnormalities \cite{zhang_mri_nephro,gervais_rcc,prasad_kidney_imaging}. All experiments in this study were performed on a clinical MRI scanner using infused urea doses which have been shown to be safe for humans with far advanced renal failure \cite{johnson_urealoading}.

\section*{Conclusion}

High resolution imaging of two key steps of the renal urea handling process was enabled by a hyperpolarized \textsuperscript{13}C relaxometry. Selective quenching of the vascular hyperpolarized \textsuperscript{13}C signal with a macromolecular relaxation agent revealed that a long-$T_2$ component of the \textsuperscript{13}C urea signal originated from the renal extravascular space, thus allowing the vascular and filtrate pools of the \textsuperscript{13}C urea to be distinguished via multi-exponential analysis. The $T_2$ response to induced diuresis and antidiuresis was performed with two imaging agents: hyperpolarized \textsuperscript{13}C urea and a control agent hyperpolarized bis-1,1-(hydroxymethyl)-1-\textsuperscript{13}C-cyclopropane-$^2\textrm{H}_8$. Large $T_2$ increases in the inner-medullar and papilla were observed with the former agent and not the latter during antidiuresis suggesting that $T_2$ relaxometry may be used to monitor the inner-medullary urea transporter (UT)-A1 and UT-A3 mediated urea concentrating process. Two high resolution imaging techniques - multiple echo time averaging and ultra-long echo time sub-2 mm$^3$ resolution 3D imaging - were developed to exploit the particularly long relaxation times observed.

\bibliographystyle{ieee}
\bibliography{greed_refs}

\end{document}